\documentstyle[11pt]{article}
\begin{document}

\title{ SUSY QCD corrections to W $\gamma$ production in polarized
        hadronic collisions
\footnote{This work was supported by National Natural Science Foundation of China}}

\author{{
 Han Meng$^{b}$, Ma Wen-Gan$^{a,b,c}$, Han Liang$^{b}$ and Jiang Yi$^{b}$}\\
{\small $^{a}$CCAST (World Laboratory), P.O.Box 8730, Beijing 100080, P.R.China}\\
{\small $^{b}$Department of Modern Physics, University of Science and Technology}\\
{\small of China (USTC), Hefei, Anhui 230027, P.R.China,}\\
{\small $^{c}$Institute of Theoretical Physics, Academia Sinica,}\\
{\small P.O.Box 2735, Beijing 100080, P.R.China. }
}

\date{}
\maketitle

\vskip 12mm
\begin{center}\begin{minipage}{5in}
\begin{center} ABSTRACT\end{center}
\baselineskip 0.3in
{We present the calculation of the supersymmetric QCD correction to
$W \gamma$ production process in polarized proton-antiproton collisions
at the TeV energy region. We find that the correction can reach
$1.2\%$ at parton level with favorable mass values of squarks and
gluino, which is comparable with the virtual correction part of
the conventional QCD.  The production rates for different polarized
$p \bar{p}$ collisions are also compared.} \\

\vskip 10mm
{~~~~PACS number(s): 13.85.Qk, 13.88.+e, 12.38.Bx, 14.80.Ly}
\end{minipage}
\end{center}

\begin{large}
\baselineskip=0.36in

\newpage
\noindent
\baselineskip 0.35in
\begin{flushleft} {\bf 1. Introduction.} \end{flushleft}

One of the important tests of the standard model(SM) for electroweak
interaction is to study the self-couplings of gauge bosons. Since the
measurement of hadronic $W^{\pm} \gamma$ production has the advantages of
very clear background, large production rate and both of the final
produced gauge bosons can be easily detected experimentally, this process
is attractive in experiment in testing the gauge boson trilinear couplings
and it is worthwhile to evaluate the W $\gamma$ production in hadronic
collisions more precisely.

$W\gamma$ production process was first investigated in
Refs.\cite{Brown}\cite{Renard}. From that time, this process has been more
carefully studied in two ways. On one hand, many theoretical papers focus
on the effective Lagrangian of $WW\gamma$ coupling \cite{Wiest}, the
related magnetic dipole moment and electric quadrapole moment of W bosons.
On the other hand, some theorists began to concentrate their attention on
the radiative corrections to this process. The calculation of
$O(\alpha_{s})$ order QCD corrections to $W\gamma$ production in hadronic
collisions in the context of the standard model(SM) was first presented in
Ref.\cite{Smith}, then was developed by J. Ohnemus \cite{Ohnemus}. The
numerical results of Ref.\cite{Ohnemus} show that the $O(\alpha_{s})$
order QCD corrections to hadronic $W\gamma$ production are significant due
to the consequence of the radiation amplitude zero(RAZ) in the Born
subprocess, it is imperative that the QCD radiative corrections should be
taken into account.

Now two questions arise naturally: Firstly, how about the radiative corrections
within other extended standard models, such as supersymmetric model(SUSY),
when the center-of-mass collision energy approaches a few TeV? Secondly,
since W boson and photon production has very low background and both particles
are easy to be identified in experiment, it enables us to conduct more precise
experiment on this process. If we can get more informations about WW $\gamma$
coupling from other experiments, such as from LEP collider, can we use this
process as an indirect probe of the new physics beyond the SM in the QCD sector?

In this paper we calculated the $O(\alpha_{s})$ SUSY QCD correction to the
process $p \bar{p} \rightarrow W^{+} \gamma + X$ in the TeV energy range,
where it is generally believed that the new physics would enter \cite{Haber}.

In the last few years there has been a resurgence of interests in the spin
structure of nucleon \cite{Ladinsky}. The polarized deep inelastic
scattering experiments(DIS) at CERN and SLAC have provided some useful
results. As to the $W\gamma$ production process at the parton level,
it is due to the V-A weak interaction between W boson and quarks, there
exits only one non-zero polarized amplitude at high energy,
i.e. $u_L+\bar{d}_R \rightarrow W^{+} + \gamma$. Consequently,
the production rates in different polarized $p \bar{p}$ collisions would
be different, which may improve our measurement in finding the SUSY QCD
signals, therefore the calculation of the $W\gamma$ production process in
polarized proton-antiproton collisions would provides an indirect probe
of the SUSY QCD.

This paper is organized as follows:  In section 2, we present the Lagrangians
\cite{Haber} concerning the interactions involved in this process and the
renormalized scheme adopted in this work\cite{Denner}.
In section 3, we give the calculation of cross section of $W\gamma$
production at the parton level. The calculation of the total one-loop
SUSY QCD corrections in the polarized hadronic process is given in
section 4. Finally the discussion and conclusion are given
in section 5.

\vskip 5mm
\begin{flushleft} {\bf 2. Theory concerned in calculation.} \end{flushleft}

The Lagrangians of the quark-squark-gluino and squark-vector-boson-squark
interactions within supersymmetric model are written explicitly as
\cite{Haber}\cite{Bartl}:

$$
\begin{array}{lll}
L_{q\tilde{q}\tilde{g}}&=&-{\sqrt{2}} g_{s} T^{\alpha}_{jk} \sum\limits^{}_{i=u,d} \left(
\bar{\tilde{g}}_{\alpha} P_{L} q^{k}_{i} \tilde{q}^{j\ast}_{iL}+
\bar{q}^{j}_{i} P_{R} \tilde{g}_{\alpha} \tilde{q}^{k}_{iL} -
\bar{\tilde{g}}_{\alpha} P_{R} q^{k}_{i} \tilde{q}^{j\ast}_{iR}-
\bar{q}^{j}_{i} P_{L} \tilde{g}_{\alpha} \tilde{q}^{k}_{iR} \right)\\

L_{\tilde{q}\tilde{q}V}&=&\frac{-ig}{\sqrt{2}}\left[
W^{+}_{\mu}(\tilde{u}^{\ast}_{L} {}^{\leftrightarrow}_{{\partial}^{\mu}}\tilde{d}_{L})+
W^{-}_{\mu}(\tilde{d}^{\ast}_{L} {}^{\leftrightarrow}_{{\partial}^{\mu}}\tilde{u}_{L})\right]
-ie A_{\mu}\sum\limits^{}_{i} e_{i}\tilde{q}^{\ast}_{i}
{}^{\leftrightarrow}_{{\partial}^{\mu}} \tilde{q}_{i}\\
\end{array}
\eqno{(2.1)}
$$

where $g_{s}$ is the strong coupling constant, $T^{a}$ is SU(3) color
generators, $q_{L}$ and $q_{R}$ are the wave functions of SU(2) weak
eigenstates.The relationship between weak eigenstates $q_{L}$ and $q_{R}$
and mass eigenstates $q_{1}$ and $q_{2}$ can be expressed as:

$$
\begin{array}{lll}
\left ( \begin{array}{l}
\tilde{q}_{L_{i}} \\ \tilde{q}_{R_{i}}
\end{array}
\right ) &=&
\left ( \begin{array}{ccc}
cos \theta_{i} e^{-i \phi} &~~ & sin \theta_{i} e^{-i \phi} \\
-sin \theta_{i} e^{i \phi} &~~ & cos \theta_{i} e^{i \phi}
\end{array} \right ) ~
\left ( \begin{array}{l}
\tilde{q}_{1,i} \\
\tilde{q}_{2,i}
\end{array} \right )
\end{array}
$$

where $\theta$ is the squark mixing angle and $\phi$ is the CP-violating
phase angle originated from the scalar quark mass matrix \cite{Han}.
In terms of mass eigenstates, the corresponding Lagrangians of
quark-squark-gluino and squark-squark-vector-boson interactions are
represented as:

$$
\begin{array} {llll}
L_{q\tilde{q}\tilde{g}}&=&-\sqrt{2} g_s T^a ~ \bar{q} [
      (A_{R} ~ \tilde{q}_1 + B_{R} ~ \tilde{q}_2) P_{R} +
      (A_{L} ~ \tilde{q}_1 - B_{L} ~ \tilde{q}_2) P_{L}] \tilde{g}_a + h.c.
\\
L_{\tilde{q}\tilde{q}V}&=&\frac{-ig}{\sqrt{2}} [
A_{R,u} B_{L,d} W^{-}_{\mu}(\tilde{d}^{\ast}_{1} {}^{\leftrightarrow}_{{\partial}^{\mu}}\tilde{u}_{1})+
A_{R,u} A_{L,d} W^{-}_{\mu}(\tilde{d}^{\ast}_{2} {}^{\leftrightarrow}_{{\partial}^{\mu}}\tilde{u}_{1})+
\\&~&~~~~~
B_{R,u} B_{L,d} W^{-}_{\mu}(\tilde{d}^{\ast}_{1} {}^{\leftrightarrow}_{{\partial}^{\mu}}\tilde{u}_{2})+
B_{R,u} A_{L,d} W^{-}_{\mu}(\tilde{d}^{\ast}_{2} {}^{\leftrightarrow}_{{\partial}^{\mu}}\tilde{u}_{2})+
\\&~&~~~~~
A_{R,d} B_{L,u} W^{+}_{\mu}(\tilde{u}^{\ast}_{1} {}^{\leftrightarrow}_{{\partial}^{\mu}}\tilde{d}_{1})+
A_{R,d} A_{L,u} W^{+}_{\mu}(\tilde{u}^{\ast}_{2} {}^{\leftrightarrow}_{{\partial}^{\mu}}\tilde{d}_{1})+
\\&~&~~~~~
B_{R,d} B_{L,u} W^{+}_{\mu}(\tilde{u}^{\ast}_{1} {}^{\leftrightarrow}_{{\partial}^{\mu}}\tilde{d}_{2})+
B_{R,d} A_{L,u} W^{+}_{\mu}(\tilde{u}^{\ast}_{2} {}^{\leftrightarrow}_{{\partial}^{\mu}}\tilde{d}_{2})]-
\\&~&~~~~~
ie A_{R,q} B_{L,q} A_{\mu} e_{i}\tilde{q}^{\ast}_{1}
{}^{\leftrightarrow}_{{\partial}^{\mu}} \tilde{q}_{1}+
B_{R,q} A_{L,q}A_{\mu} e_{i}\tilde{q}^{\ast}_{2}
{}^{\leftrightarrow}_{{\partial}^{\mu}} \tilde{q}_{2}
\end{array}
\eqno{(2.2)}
$$
Where
$$
\begin{array}{lll}
A_{L,i} = cos \theta_i e^{-i \phi}~~~~B_{L,i} &=& sin \theta_{i} e^{-i \phi}\\
A_{R,i} = sin \theta_i e^{i \phi}~~~~~~ B_{R,i} &=& cos \theta_{i} e^{i \phi}
\end{array}
$$

In order to eliminate the ultraviolet divergences that appear in the one-loop
integrals, we adopt the on-mass-shell(OMS) and dimensional regularization
schemes \cite{Denner}. The renormalized irreducible two point functions
for fermions are defined as\cite{Han}:
\begin{eqnarray*}
&~&~~~~~\hat{\Gamma}(p) = i (\rlap/p-m_{t}) + i \left [ \rlap/p P_{L}
                         \hat{\Sigma}^{L}(p^2)
                        + \rlap/p P_{R} \hat{\Sigma}^{R}(p^2)
                        + P_{L} \hat{\Sigma}^{S}(p^2)
                        + P_{R} \hat{\Sigma}^{S~\ddag}(p^2) \right].\\
&~&\rm{From ~the~ renormalization~ conditions~ for~ the~ on-mass-shell ~physical ~field \cite{Denner}:}\\
&~&~~~~~\tilde{Re}\bar{u_q}\hat{\Gamma}_{q}(p)\vert_{p^2=m^2_q}=0\\
&~&~~~~~\lim_{p^{2}\rightarrow m_q^2 }~~ \bar{u}_q(p) \tilde{Re}~~
\hat{\Gamma}_{q}(p)\frac{\rlap/{p}+m_{q}}{p^2-m^{2}_{q}}=i\bar{u_{q}}(p),\\
&~&\rm{~we ~can~ deduce~ all~ the~ related~ renormalization~ constants ~involved ~in ~this}\\
&~&\rm{~process \cite{Han}~ as:}\\
&~&~~~~~\delta m_q= \frac{m_q}{2} \tilde{Re} \left(
\Sigma^{L}(m_q^2) + \Sigma^{R}(m_q^2) + \Sigma^{S}(m_q^2) + \Sigma^{S\ddag}(m_q^2)\right)\\
&~&~~~~~\delta Z^{L}= -\tilde{Re} \Sigma^{L}(m_q^2) - m_q^2 \frac{\partial}{\partial{p^2}}
\left[ \tilde{Re} \left( \Sigma^{L}(p^2) + \Sigma^{S}(p^2) + \Sigma^{S\ddag}(p^2) + \Sigma^{R}(p^2)
\right) \right]\\
&~&~~~~~\delta Z^{R}= -\tilde{Re} \Sigma^{R}(m_q^2) - m_q^2 \frac{\partial}{\partial{p^2}}
\left[ \tilde{Re} \left( \Sigma^{L}(p^2) + \Sigma^{S}(p^2) + \Sigma^{S\ddag}(p^2)+\Sigma^{R}(p^2)
\right) \right] +\\
&~&~~~~~~~~~~\frac{1}{m} (\Sigma^{S\ddag}(m^2)-\Sigma^{S}(m^2))
\end{eqnarray*}

One can refer to Ref. \cite{Han} for the relevant unrenormalized self-energies
and the counterterms take the following forms:

$$
\begin{array} {lll}
i \delta \Sigma &=& i [
  C_{L} \rlap/p P_{L} + C_{R} \rlap/p P_{R} - C^{-}_{S} P_{L} - C^{+}_{S} P_{R}
  ], \\
  \\
i \delta \Lambda^{\mu} &=& -i e Q_e \gamma^{\mu} [
C^{-} P_{L} + C^{+} P_{R} ]. \\
\end{array}
$$

$$
\begin{array} {lll}
C_L &=& C^{-} = \frac{1}{2} (\delta Z^L + \delta Z^{L\dag}), \\
C_R &=& C^{+} = \frac{1}{2} (\delta Z^R + \delta Z^{R\dag}), \\
C^{-}_{S} &=& \frac{m_{q}}{2} (\delta Z^L + \delta Z^{R\dag}) +
              \delta m_{q}, \\
C^{+}_{S} &=& \frac{m_{q}}{2} (\delta Z^R + \delta Z^{L\dag}) +
              \delta m_{q}.
\end{array}
\eqno{(2.3)}
$$

\vskip 5mm
\begin{flushleft} {\bf 3. The calculation of the subprocess
            $u \bar{d} \rightarrow W^{+} \gamma$.} \end{flushleft}

We denote the subprocess which we calculated as
$$
u (p_3, L) \bar{d} (p_4, R) \longrightarrow W^{+}(p_1) \gamma(p_2).
\eqno{(3.1)}
$$
The $O(\alpha_{s})$ order SUSY QCD correction to (3.1) comes from the
interference between the tree-level graphs shown in Fig.1 and the
one-loop graphs shown in Fig.2. The renormalized amplitude for
$u (p_3)_{L} \bar{d} (p_4)_{R} \longrightarrow W^{+}(p_1) \gamma(p_2)$
including the $O(\alpha_{s})$ SUSY QCD corrections can be written as:
$$
\begin{array} {lll}
\vert M^{ren}_{L,R} \vert^{2}&=& \vert M^{tree}_{L,R}  \vert^{2}+
               2~ Re~ (M^{tree \dagger}_{L,R}  \delta{M}^{1-loop}_{L,R} )\\
\delta M^{1-loop}_{L,R} &=& \delta{M}^{self}_{L,R} + \delta{M}^{tri}_{L,R} +
       \delta{M}^{box}_{L,R}.
\end{array}
\eqno{(3.2)} $$

$M_{L,R}^{tree}$ is the amplitude of the tree level. $\delta{M}_{L,R}^{self}$,
$\delta{M}_{L,R}^{tri}$ and $\delta{M}_{L,R}^{box}$ represent the renormalized
amplitudes coming from the self-energy, triangle and box diagrams,
respectively. The lower indices ${L,R}$ in above amplitude notations
represent the matrix elements for the process $u_{L} (p_3) \bar{d}_{R} (p_4)
\longrightarrow W^{+}(p_1) \gamma(p_2) $, which is the only non-zero
contribution process in all polarized parton cases. The one-loop and Born
amplitude can be written as:

\begin{eqnarray*}
\delta M^{self}_{L,R} &=& \delta{M}^{self}_{\tilde{u}\tilde{u}}+
                          \delta{M}^{self}_{\tilde{d}\tilde{d}} \\
\delta M^{tri}_{L,R} &=&  \delta{M}^{tri}_{\tilde{u}\tilde{u}\gamma}+
                          \delta{M}^{tri}_{\tilde{d}\tilde{d}\gamma}+
                          \delta{M}^{tri,1}_{\tilde{u}\tilde{d} w}+
                          \delta{M}^{tri,2}_{\tilde{u}\tilde{d} w}+
                          \delta{M}^{tri,3}_{\tilde{u}\tilde{d} w}\\
M^{tree}&=& \frac{e g}{\sqrt{2}}
        \epsilon_{\mu}(p_2)\epsilon_{\nu}(p_1) \bar{v}_{R}(p_3)\\
        &~& \left\{ \frac{2}{3 u}\gamma^{\nu} (\rlap/p_{4} - \rlap/p_{2}) \gamma^{\mu}-
        \frac{1}{3 t}\gamma^{\mu} (\rlap/p_{3} - \rlap/p_{1}) \gamma^{\nu}- \right. \\
        &~&\left. \frac{1}{s-m_W^2}[(\rlap/p_1-\rlap/p_2) g^{\mu\nu}+
        2 p_2^{\nu} \gamma^{\mu} - 2 p_1^{\mu} \gamma^{\nu}] \right\} u_{L}(p_4)
        ~~~~~~~(3.3)
\end{eqnarray*}

The amplitudes for different groups of one-loop diagrams can be expressed
separately. That means we denote $\delta M^{self}_{\tilde{u}\tilde{u}},
\delta M^{self}_{\tilde{d}\tilde{d}}$ as the amplitudes which correspond
to Fig.2(9) and Fig.2(10) respectively. For the triangle diagrams, $\delta
M^{tri}_{\tilde{u}\tilde{u}\gamma},
\delta M^{tri,1}_{\tilde{u}\tilde{d}w},
\delta M^{tri}_{\tilde{d}\tilde{d}\gamma},
\delta M^{tri,2}_{\tilde{u}\tilde{d}w},
\delta M^{tri,3}_{\tilde{u}\tilde{d}w} \rm{and}~
\delta M^{tri}_{\tilde{u}\tilde{d}w\gamma}$
correspond to amplitudes of Fig.2(1), Fig.2(2), Fig.2(3), Fig.2(4), Fig.2(7),
Fig.2(8), respectively.
$\delta M^{box,1}$ and $\delta M^{box,2} $ represent the contributions
from box diagrams Fig.2(5) and Fig.2(6), respectively.
The explicit expressions for all the amplitude parts involving in the one-loop
amplitude $\delta{M}$ are listed in Appendix.

Then the cross section for subprocess $u_{L} (p_3) \bar{d}_{R} (p_4)
\longrightarrow W^{+}(p_1) \gamma(p_2)$ is given by

$$
\hat{\sigma}_{L,R}(\hat{s})=\frac{1}{4}\frac{1}{9}
             \frac{ N_{c} (\hat{s}-m_{W}^2) }
             {32 \pi \hat{s}^2 }
             \int d(\cos{\theta}) \vert M^{ren}_{L,R}(\hat{s}) \vert^2,
\eqno{(3.4)}
$$

In equation (3.3) $N_{c}$ is the number of colors. The factors
$\frac{1}{4}$ and $\frac{1}{9}$ are the spin and color averages
respectively.

\vskip 5mm
\begin{flushleft} {\bf 4. The SUSY QCD corrections to $W \gamma$ production in
polarized hadronic collisions.} \end{flushleft}

Now we consider the $W\gamma$ production in hadronic collision. The process
is represented as:
$$
p (P_3,\lambda) \bar{p} (P_4,\lambda) \rightarrow u \bar{d}
\rightarrow W^{+}(p_1) \gamma(p_2) + X,
$$
where $\lambda$ denotes the helicities of the initial
proton and anti-proton. $P_3$ and $P_4$ are the four momenta of the proton
and anti-proton respectively. The total cross section for the production
of W boson and photon in polarized hadronic collisions can be written in
the form as:

$$
\begin{array}{lll}
\sigma_{p,\bar{p}}(s,\lambda)&=&\int dx_{1}\int dx_{2} \hat{\sigma}_{LR} (\hat{s})
                 [u_{\pm}(x_{1}) d_{\mp}(x_{2})]\\\\
\end{array}
\eqno{(4.1)}
$$

where the cross section at parton level $\hat{\sigma}_{LR}$ is for the
subprocess of left-handed up-quark and right-handed anti-down-quark
collisions, i.e. $u_{L}+\bar{d}_{R} \longrightarrow W + \gamma $.
$u_{\pm}(x)$ and $d_{\mp}(x)$ are the polarized parton distribution
functions. The upper signs of the indices of $u_{\pm}(x)$ and
$d_{\mp}(x)$ are for $\lambda=L$ and the lower signs are for $\lambda=R$.

$$
\begin{array} {lll}
&s=(P_3+P_4)^2\\
&\hat{s}=(p_3+p_4)^2 \\
&\hat{s}=x_{1} x_{2} s\\
&p_{j}=x_{j} P_{i} ~~~~~\rm(j=3,4)
\end{array}
$$

where $x_{1}$ and $x_{2}$ are the momentum fractions of initial
partons. For any type of partons we take the definitions as\cite{Wiest}

$$
\begin{array} {lll}
f_{+}(x) &=& \frac{1}{2} (f(x) + \Delta f(x))\\
  &=& density~ of ~L(or R) ~parton ~in~  L(or R) ~proton\\\\
f_{-}(x) &=& \frac{1}{2} (f(x) - \Delta f(x))\\
  &=& density~ of ~L(or R) ~parton ~in~  R(or L) ~proton
\end{array}
\eqno{(4.2)}
$$

In Equation(4.2) f~=~u~,d~,s, and x is the momentum fraction of the parton.

In the past few years, there are a number of parameterizations
for the polarized parton distribution functions(PDF) as shown in literature
\cite{Ladinsky}. It was indicated in this paper\cite{Ladinsky} that the
PDF's for up and down quarks are the best defined by the present experimental
data and there is nice agreement for $\Delta u$ for different parameterizations
of PDF, when x is in the range of $10^{-3}$ to $10^{-1}$. In our calculation
we adopted Brodsky parameterization for polarized quark distribution functions
as a numerical example. Brodsky's parton distribution functions can be
expressed as\cite{BBS}

$$
\begin{array} {cc} u_{+}(x) = \frac{1}{x^{\alpha}} [A_{u} (1-x)^{3} +
B_{u} (1-x)^{4}]\\ d_{+}(x) = \frac{1}{x^{\alpha}} [A_{d} (1-x)^{3} +
B_{d} (1-x)^{4}]\\\\ u_{-}(x) = \frac{1}{x^{\alpha}} [C_{u} (1-x)^{5} +
D_{u} (1-x)^{6}]\\ d_{-}(x) = \frac{1}{x^{\alpha}} [C_{d} (1-x)^{5} +
D_{d} (1-x)^{6}]\\\\ s_{+}(x) = \frac{1}{x^{\alpha}} [A_{s} (1-x)^{5} +
B_{s} (1-x)^{6}]\\ s_{-}(x) = \frac{1}{x^{\alpha}} [C_{s} (1-x)^{7} +
D_{s} (1-x)^{6}]\\ \end{array} \eqno{(4.3)}
$$

In Brodsky's(BBS) parameterization scheme, the leading Regge behavior
at $x \rightarrow 0$ has the intercept $\alpha=1.12$. By choosing this
value of $\alpha$, it allowed a good match to the unpolarized PDF given
by MRS\cite{MRS}, if we average the polarized densities. In addition, to be
consistent with the sum rules and dynamical constraint indicated in
Ref\cite{BBS}, the other eight parameters are set to be:

$$
\begin{array}{lll}
A_{u}&=&3.784,~~~~~~~~A_{d}=0.775,~~~~~~~~~A_{s}=0.2897\\
B_{u}&=&-3.672,~~~~~~B_{d}=-0.645,~~~~~~B_{s}=-0.2637\\
C_{u}&=&2.004,~~~~~~~~C_{d}=3.230,~~~~~~~~~C_{s}=1.0\\
D_{u}&=&-1.892,~~~~~~D_{d}=-3.118,~~~~~~D_{s}=-0.9725\\
\end{array}
$$

\vskip 5mm
\begin{flushleft} {\bf 5. Numerical calculation and discussion.} \end{flushleft}

In the numerical evaluation we take the input parameters as $m_{W}=
80.226~ GeV, m_{Z}=91.1887~ GeV, G_{F}=1.166392 \cdot 10^{-5} (GeV)^{-2}$ and
$\alpha=\frac{1}{137.036}$. The strong coupling constant $\alpha_{s}$ is
determined by

$$
\alpha_{s}(\mu)= \frac{\alpha_{s}(m_Z)}{1+\frac{33-2 N_f}{6 \pi} \alpha_s(m_Z)
\ln{\frac{\mu}{m_Z}}},
$$

where $\alpha_s(m_Z)=0.117$ and $N_f$ is the number of active flavors at
energy scale $\mu$. We set the transverse momentum cut of photon as
$p_{T}(\gamma) > 20 ~GeV$. This is a typical experiment acceptance cut
value. This cut is also necessary in regulating the collinear divergence
associated with the photon. Since here we don't consider the CP-violation
effect in this process, we set the CP-violation angel $\phi$ to be zero.
In order to illuminate the SUSY QCD effects more clearly and for the
sake of simplicity, we assume that the scalar quark masses are all
degenerate, i.e. $m_{\tilde{u_1}}$~=~$m_{\tilde{u_2}}$~=~
$m_{\tilde{d_1}}$~=~$m_{\tilde{d_2}}$ ~=~$m_{\tilde{q}}$.

If we take the experimental mass bounds on squarks and gluino into account,
the corresponding mass parameters are set to be $m_{\tilde{q}} > 175 ~GeV$
(for $m_{\tilde{g}} < 300 GeV$ ) and $m_{\tilde{g}} > 175 GeV$ \cite{Beenakker}.
Since the very light gluino mass is not excluded experimentally and there
has been renewed interest in this case recently, we also set the gluino mass
to be $5~ GeV$ as a comparison.

In Fig.3 we present the relative $O(\alpha_{s})$ SUSY QCD corrections
($\delta=\frac{\Delta \sigma}{\sigma_{tree}}$) to
the $ W \gamma $ hadronic production at the parton level. The two curves
correspond to two different values of squark and gluino masses, namely, solid
curve for $m_{\tilde{g}}=5~GeV, m_{\tilde{q}}=175~GeV$ and dashed curve for
$m_{\tilde{g}}=150 GeV, m_{\tilde{q}}=175~GeV$. From Fig.3 we can see that
the relative corrections can reach $1.2\%$ with $m_{\tilde{g}}= 5~GeV$ when
$\sqrt{\hat{s}}=500~GeV $.

The relationships between the SUSY QCD corrections and masses of the
squark and gluino are illustrated in Fig.4 and Fig.5 respectively, where
$\sqrt{\hat{s}}=500 GeV$. We find that the corrections decrease heavily
with the increasing of $m_{\tilde{q}}$ and $m_{\tilde{g}}$, they show
clearly the effects of decoupling.

In our calculation we find that the contribution from the box diagrams and
the quartic vertex diagrams are much less than those from other triangle
and self-energy diagrams. As the result, the resonance effect at
$\sqrt{\hat{s}}=2~m_{\tilde{q}}$ is suppressed, so there is no obvious
peak in Fig.3.

Fig.6 shows the relative SUSY QCD corrections to $W \gamma$ production in
the polarized proton-antiproton collisions, where the solid curve
corresponds to $p_{L} \bar{p}_{L}$ collision and the dashed curve
represents $p_{R} \bar{p}_{R}$ collision. The relative correction
discrepancy between these two different polarized $p \bar{p}$ collisions
is obvious and the absolute value of relative corrections increases with
the increment of $\sqrt{s}$. The relative correction for $p_{L}
\bar{p}_{L}$ collision at $\sqrt{s}=2~TeV$ reaches $0.2\%$.

The calculation in Ref.\cite{Ohnemus} states that the SM QCD virtual
relative correction is only about $1 \%$ of the Born approximation,
whereas our numerical results show that the corresponding SUSY QCD virtual
relative corrections could reach $1.2\%$ with favorable mass parameters,
which is of the same order of the SM QCD virtual corrections.

As we know, in $u \bar{d} \rightarrow W \gamma $ subprocess only the first
generation of the scalar quarks contributes to the SUSY QCD correction. In
order to illuminate the SUSY QCD effects, we assumed that the masses of
three generations of the scalar quarks are degenerate in our calculation.
However, if the degeneracy of the scalar quarks was lifed, it is generally
believed that the $\tilde{u}$, $\tilde{d}$ would be the heaviest squarks,
the SUSY signals would be weaker due to the decoupling effect.

In practical experiment, the soft gluon radiation, hard collinear
corrections etc. should be also included in the conventional QCD
corrections. Their combined contributions could result in large
corrections to $p + \bar{p} \rightarrow W + \gamma + X $ inclusive
process, which would be strong background to the SUSY signal. However, if
we impose the jet veto in experiment, the size of the SM QCD corrections
would reduce to only about few percent of the Born approximation in the $p
+ \bar{p} \rightarrow W \gamma + 0~jet$ process at the Tevatron
energy\cite{Ohnemus}. Since both supersymmetric and conventional QCD
corrections are mainly negative in the 0-jet process, it could be possible
to disentangle the SUSY signals from the conventional QCD effects by more
precise measurement of the $p + \bar{p} \rightarrow W \gamma$ process.

At high colliding energy, u and d quarks are approximately massless
particles, the W boson couples only to left-handed u quark and
right-handed $\bar{d}$ quark. And we know that the densities of
left-handed and right-handed quarks are not equal for polarized proton.
Therefore the cross section for hadronic W $\gamma$ production should be
of significant difference for different polarized condition. Fig.6 shows
that the relative SUSY QCD corrections for $p_L \bar{p}_L$ collision are
larger than the $p_R \bar{p}_R$ collisions, it could be used to improve
the experimental measurement for SUSY corrections.

In this paper we evaluated the SUSY QCD corrections to the $p \bar{p}
\rightarrow u \bar{d} \rightarrow W^{+} \gamma + X $ process and presented
the complete analytic expressions and numerical results including the one-loop
SUSY QCD virtual corrections. The relative corrections can reach $1.2\%$ for
$m_{\tilde{g}}=5 GeV$, $m_{\tilde{u}}=m_{\tilde{d}}=175 GeV$ at the parton
level and about $0.2\%$ after convoluted with the parton distribution functions.
The corrections are sensitive to the masses of the squarks and gluino,
especially for very light gluino. We can conclude that the SUSY QCD
corrections with appropriate superpartner masses for the process
$p \bar{p} \rightarrow u \bar{d} \rightarrow W^{+} \gamma + X$ can be
noticeful on TeV energy scale.

\begin{flushleft} {\bf Acknowledgement} \end{flushleft}
We are very grateful to Z.H. Yu for his many stimulating discussions.
Thanks also go to M.L. Zhou and G. Lin for their help in drawing the
Feynman diagrams.

\begin{flushleft} {\bf Appendix: One-loop amplitudes and form factors} \end{flushleft}

\begin{eqnarray*}
&~&\rm{The~one-loop~amplitude~parts~appearing~in Eq.(3.3)~are~expressed~ as:}\\
\delta M^{self}_{\tilde{u}\tilde{u}}  &=&\frac{-i}{u^2} \frac{ e g g^{2}_{s}} {3 \sqrt{2}} \frac{C_{f}}{4 \pi^{2}}
\epsilon_{\mu}(p_{2}) \epsilon_{\nu} (p_{1}) \bar{v}_{R}(p_{3}) \\
        &~&(
        f^{\tilde{u}\tilde{u}}_{1} \gamma^{\nu} \rlap/p_{4} \rlap/p_{2}  \gamma^{\mu} +
        2 f^{\tilde{u}\tilde{u}}_{2}  p_{2} \cdot p_{4} \gamma^{\nu} \gamma^{\mu} +
        4 f^{\tilde{u}\tilde{u}}_{3} \gamma^{\nu} p^{\mu}_{4} +
        2 f^{\tilde{u}\tilde{u}}_{4} \gamma^{\nu} \rlap/p_{2} p^{\mu} _{4}
        ) u_{L}(p_{4})
\\
\delta M^{self}_{\tilde{d}\tilde{d}}  &=& \frac{i}{t^2}
       \frac{e g g^{2}_{s}} {3 \sqrt{2}} \frac{C_{f}}{8 \pi^{2}}
\epsilon_{\mu}(p_{2}) \epsilon_{\nu} (p_{1}) \bar{v}_{R}(p_{3})
        (
        2 f^{\tilde{d}\tilde{d}}_{1}  p_{3} \cdot p_{2} \gamma^{\mu} \gamma^{\nu} +
        4 f^{\tilde{d}\tilde{d}}_{2} \gamma^{\nu} p^{\mu}_{3}
        ) u_{L}(p_{4})
\\
\delta M^{tri}_{\tilde{u}\tilde{u}\gamma}  &=& \frac{-i}{u} \frac{ e g g^{2}_{s}} {3 \sqrt{2}} \frac{C_f}{4 \pi^{2}}
\epsilon_{\mu}(p_{2}) \epsilon_{\nu} (p_{1}) \bar{v}_{R}(p_{3})\\
        &~&(
        f^{\tilde{u}\tilde{u}\gamma}_{1} \gamma^{\nu} \rlap/p_{2} \gamma^{\mu} +
        2 f^{\tilde{u}\tilde{u}\gamma}_{2} \gamma^{\nu} \rlap/p_{2} p^{\mu}_{4} +
        2 f^{\tilde{u}\tilde{u}\gamma}_{3} \gamma^{\nu} p^{\mu}_{4}
        ) u_{L}(p_{4})
\\
\delta M^{tri}_{\tilde{d}\tilde{d}\gamma} &=&\frac{i}{t} \frac{e g g^{2}_{s}} {3 \sqrt{2}} \frac{C_{f}}{8 \pi^{2}}
\epsilon_{\mu}(p_{2}) \epsilon_{\nu} (p_{1}) \bar{v}_{R}(p_{3})
        (
        f^{\tilde{d}\tilde{d}}_{1} \gamma^{\mu} \rlap/p_{2} \gamma^{\nu} +
        2 f^{\tilde{d}\tilde{d}}_{2} \gamma^{\nu} p^{\mu}_{3}
        ) u_{L}(p_{4})
\\
\delta M^{tri,1}_{\tilde{u}\tilde{d} w} &=&\frac{i}{t} \frac{ e g g^{2}_{s}} {3 \sqrt{2}} \frac{C_{f}}{4 \pi^{2}}
\epsilon_{\mu}(p_{2}) \epsilon_{\nu} (p_{1}) \bar{v}_{R}(p_{3})\\
        &~&(
        \frac{1}{4} f^{\tilde{u}\tilde{d} w,1}_{1} \gamma^{\mu} \rlap/p_{2} \gamma^{\nu} +
        f^{\tilde{u}\tilde{d} w,1}_{2} \gamma^{\mu} \rlap/p_{2} p^{\nu}_{4}  +
        f^{\tilde{u}\tilde{d} w,1}_{3} \gamma^{\mu} \rlap/p_{2} \rlap/p_{1} p^{\nu}_{4} +\\
        &~&
        \frac{1}{2} f^{\tilde{u}\tilde{d} w,1}_{4} \gamma^{\nu} p^{\mu}_{3} +
        2 f^{\tilde{u}\tilde{d} w,1}_{5} p^{\mu}_{3} p^{\nu}_{4} +
        2 f^{\tilde{u}\tilde{d} w,1}_{6} \rlap/p_{1} p^{\mu}_{3} p^{\nu}_{4}) u_{L}(p_{4})
\\
\delta M^{tri,2}_{\tilde{u}\tilde{d} w} &=&\frac{-i}{u} \frac{\sqrt{2} e g g^{2}_{s}}{3} \frac{C_{f}}{8 \pi^{2}}
\epsilon_{\mu}(p_{2}) \epsilon_{\nu} (p_{1}) \bar{v}_{R}(p_{3})\\
        &~&(
        \frac{1}{2} f^{\tilde{u}\tilde{d} w,2}_{1} \gamma^{\nu} \rlap/p_{2} \gamma^{\mu} +
        2 f^{\tilde{u}\tilde{d} w,2}_{2} \rlap/p_{2} \gamma^{\mu} p^{\nu}_{3} +
        2 f^{\tilde{u}\tilde{d} w,2}_{3} \rlap/p_{1} \rlap/p_{2} \gamma^{\mu} p^{\nu}_{3} +\\
        &~&
        f^{\tilde{u}\tilde{d} w,2}_{4} \gamma^{\nu} p^{\mu}_{4} +
        4 f^{\tilde{u}\tilde{d} w,2}_{5} p^{\nu}_{3} p^{\mu}_{4}+
        4 f^{\tilde{u}\tilde{d} w,2}_{6} \rlap/p_{1} p^{\nu}_{3} p^{\mu}_{4}
\\
\delta M^{tri,3}_{\tilde{u}\tilde{d} w} &=&\frac{i}{s} \frac{ \sqrt{2}
e g g^{2}_{s} C_f} {8 \pi^{2}}
\epsilon_{\mu}(p_{2}) \epsilon_{\nu} (p_{1}) \bar{v}_{R}(p_{3})\\
&~&(f^{\tilde{u}\tilde{d} w,3}_{1} g^{\mu\nu}  \rlap/p_{1}  +
f^{\tilde{u}\tilde{d} w,3}_{2} g^{\mu\nu}  \rlap/p_{2}  +
f^{\tilde{u}\tilde{d} w,3}_{3}   \gamma^{\nu}   p^{\mu}_{1}+
f^{\tilde{u}\tilde{d} w,3}_{4}   \gamma^{\mu}  p^{\nu}_{2}+\\
&~& f^{\tilde{u}\tilde{d} w,3}_{5}   p^{\nu}_{2}p^{\mu}_{4}+
f^{\tilde{u}\tilde{d} w,3}_{6}   \rlap/p_{1}  p^{\nu}_{2}p^{\mu}_{4}+
f^{\tilde{u}\tilde{d} w,3}_{7}   \rlap/p_{2}  p^{\nu}_{2}p^{\mu}_{4}+
f^{\tilde{u}\tilde{d} w,3}_{8}   p^{\mu}_{1}p^{\nu}_{4}+\\
&~&f^{\tilde{u}\tilde{d} w,3}_{9}   \rlap/p_{1}  p^{\mu}_{1}p^{\nu}_{4}+
f^{\tilde{u}\tilde{d} w,3}_{10}   \rlap/p_{2}  p^{\mu}_{1}p^{\nu}_{4}
) u_{L}(p_{4})\\
\delta M^{tri}_{\tilde{u}\tilde{d}w\gamma}&=& \frac{i e g g^2_s C_f} {3 \sqrt{2}}
\frac{1} {8 \pi^{2}} \epsilon_{\mu}(p_{2}) \epsilon_{\nu}
(p_{1}) \bar{v}_{R}(p_{3}) f^{\tilde{u}\tilde{d} w \gamma} g^{\mu\nu}
u_{L}(p_{4})\\
\delta M^{box,1}_{L,R}&=&
\frac{i e g g^2_s }{3 \sqrt{2}} \frac{C_f}{2 \pi^2} \epsilon_{\mu}(p_{2}) \epsilon_{\nu} (p_{1})\bar{v}_{R}(p_3)
( f^{box,1}_{1} g^{\mu\nu}+
f^{box,1}_{2}g^{\mu\nu} \rlap/p_2+ f^{box,1}_{3}\gamma^{\nu} p^{\mu}_1+
f^{box,1}_{4}\gamma^{\mu} p^{\nu}_2+f^{box,1}_{5} p^{\mu}_1 p^{\nu}_2+\\
&~&f^{box,1}_{6}\rlap/p_2 p^{\mu}_1 p^{\nu}_2+ f^{box,1}_{7}\gamma^{\nu} p^{\mu}_3+
f^{box,1}_{8}p^{\nu}_2 p^{\mu}_3+ f^{box,1}_{9}\rlap/p_2p^{\nu}_2 p^{\mu}_3+\\
&~&f^{box,1}_{10}\gamma^{\mu} p^{\nu}_3+ f^{box,1}_{11}p^{\mu}_1 p^{\nu}_3+
f^{box,1}_{12} \rlap/p_2 p^{\mu}_1 p^{\nu}_3+ f^{box,1}_{13}p^{\mu}_3 p^{\nu}_3+\\
&~&f^{box,1}_{14} \rlap/p_2 p^{\mu}_3 p^{\nu}_3 ) u_{L}(p_4)\\
\delta M^{box,2}_{L,R}&=&
\frac{i e g g^2_s }{3 \sqrt{2}} \frac{C_f}{2 \pi^2} \epsilon_{\mu}(p_{2}) \epsilon_{\nu} (p_{1}) \bar{v}_{R}(p_3)
(f^{box,2}_{1} g^{\mu\nu}+
f^{box,2}_{2}g^{\mu\nu} \rlap/p_1+ f^{box,2}_{3}\gamma^{\nu} p^{\mu}_1+
f^{box,2}_{4}\gamma^{\mu} p^{\nu}_2+f^{box,2}_{5} p^{\mu}_1 p^{\nu}_2+\\
&~&f^{box,2}_{6}\rlap/p_1p^{\mu}_1 p^{\nu}_2+ f^{box,2}_{7}\gamma^{\nu} p^{\mu}_3+
f^{box,2}_{8}p^{\nu}_2 p^{\mu}_3+ f^{box,2}_{9}\rlap/p_1p^{\nu}_2 p^{\mu}_3+\\
&~&f^{box,2}_{10}\gamma^{\mu} p^{\nu}_3+ f^{box,2}_{11}p^{\mu}_1 p^{\nu}_3+
f^{box,2}_{12}\rlap/p_1p^{\mu}_1 p^{\nu}_3+ f^{box,2}_{13}p^{\mu}_3 p^{\nu}_3+\\
&~&f^{box,2}_{14} \rlap/p_1 p^{\mu}_3 p^{\nu}_3) u_{L} (p_4)\\
\end{eqnarray*}

In above equations $C_{f}=\frac{4}{3}$ which is the color factor arising
from the quark-squark-gluino vertices in one-loop diagrams.
The definition of one-loop integral functions are adopted from Ref.\cite{Bernd}.

\begin{eqnarray*}
f^{\tilde{u}\tilde{u}}_{1}=&~&
    - cos \theta sin \theta ( B_0[m_{u}, m_{\tilde{g}}, m_{\tilde{u_1}}] +
      B_0[m_{u}, m_{\tilde{g}}, m_{\tilde{u_2}}] )+\\
&~&   cos \theta sin \theta  (B_0[p_{2} - p_{4}, m_{\tilde{g}}, m_{\tilde{u_1}}] -
      B_0[p_{2} - p_{4}, m_{\tilde{g}}, m_{\tilde{u_2}}])\\
f^{\tilde{u}\tilde{u}}_{2}=&~&
      - cos^2 \theta B_1[m_{u}, m_{\tilde{g}}, m_{\tilde{u_1}}] -
      sin^2 \theta B_1[m_{u}, m_{\tilde{g}}, m_{\tilde{u_2}}] -\\
&~&   cos^2 \theta B_1[p_{2} - p_{4}, m_{\tilde{g}}, m_{\tilde{u_1}}] -
      sin^2 \theta B_1[p_{2} - p_{4}, m_{\tilde{g}}, m_{\tilde{u_2}}]\\
f^{\tilde{u}\tilde{u}}_{3}=&~&
      p_2 \cdot p_4~ cos^2 \theta B_1[m_{u}, m_{\tilde{g}}, m_{\tilde{u_1}}] +
      p_2 \cdot p_4~ sin^2 \theta B_1[m_{u}, m_{\tilde{g}}, m_{\tilde{u_2}}] -\\
&~&   p_2 \cdot p_4~ cos^2 \theta B_1[p_{2} - p_{4}, m_{\tilde{g}}, m_{\tilde{u_1}}] -
      p_2 \cdot p_4~ sin^2 \theta B_1[p_{2} - p_{4}, m_{\tilde{g}}, m_{\tilde{u_2}}]\\
f^{\tilde{u}\tilde{u}}_{4}=&~&
    - cos \theta sin \theta  ( B_0[m_{d}, m_{\tilde{g}}, m_{\tilde{u_{1}}}] +
      B_0[m_{d}, m_{\tilde{g}}, m_{\tilde{u_{2}}}]) +\\
&~&   cos \theta sin \theta ( B_0[p_{2} - p_{4}, m_{\tilde{g}}, m_{\tilde{u_{1}}}] -
      B_0[p_{2} - p_{4}, m_{\tilde{g}}, m_{\tilde{u_{2}}}] )\\
f^{\tilde{d}\tilde{d}}_{1}=&~&
     cos^2 \theta  B_1[m_{d}, m_{\tilde{g}}, m_{\tilde{d_1}}] +
     sin^2 \theta  B_1[m_{d}, m_{\tilde{g}}, m_{\tilde{d_2}}] -\\
&~&  cos^2 \theta B_1[-p_{2} + p_{3}, m_{\tilde{g}}, m_{\tilde{d_1}}] -
     sin^2 \theta B_1[-p_{2} + p_{3}, m_{\tilde{g}}, m_{\tilde{d_2}}]\\
f^{\tilde{d}\tilde{d}}_{2}=&~&
     - p_2 \cdot p_3 cos^2 \theta  B_1[m_{d}, m_{\tilde{g}}, m_{\tilde{d_1}}] -
     p_2 \cdot p_3 sin^2 \theta  B_1[m_{d}, m_{\tilde{g}}, m_{\tilde{d_2}}] +\\
&~&  p_2 \cdot p_3 cos^2 \theta B_1[-p_{2} + p_{3}, m_{\tilde{g}}, m_{\tilde{d_1}}] +
     p_2 \cdot p_3 sin^2 \theta B_1[-p_{2} + p_{3}, m_{\tilde{g}}, m_{\tilde{d_2}}]
\end{eqnarray*}

For simplicity, we give denotations to represent the complete expressions of
C,D integral functions at first, then the lengthy arguments of C, D functions
can be omitted.

\begin{eqnarray*}
\{  C_{i}^{(1)},C_{ij}^{(1)}  \}&=&\{ C_{i}^{(1)},C_{ij}^{(1)} \}[-p_4, p_2, m_{\tilde{g}}, m_{\tilde{u_{1}}}, m_{\tilde{u_{1}}}]\\
\{  C_{i}^{(2)},C_{ij}^{(2)}  \}&=&\{ C_{i}^{(2)},C_{ij}^{(2)} \}[-p_4, p_2, m_{\tilde{g}}, m_{\tilde{u_{2}}}, m_{\tilde{u_{2}}}]\\
\end{eqnarray*}
\begin{eqnarray*}
f^{\tilde{u}\tilde{u}\gamma}_{1}=&~&
  -  sin^2 \theta  B_1[m_{u}, m_{\tilde{g}}, m_{\tilde{u_1}}] -
     cos^2 \theta  B_1[m_{u}, m_{\tilde{g}}, m_{\tilde{u_2}}] -\\
&~&   2 cos^2 \theta C_{24}^{(1)}- 2  sin^2 \theta C_{24}^{(2)}\\
f^{\tilde{u}\tilde{u}\gamma}_{2}=&~&
   - cos \theta sin \theta C_0+ cos \theta sin \theta C_0+
     cos \theta sin \theta C_{11}^{(1)}- cos \theta sin \theta C_{11}^{(2)}\\
f^{\tilde{u}\tilde{u}\gamma}_{3}=&~&
     sin^2 \theta  B_1[m_{u}, m_{\tilde{g}}, m_{\tilde{u_1}}] +
     cos^2 \theta  B_1[m_{u}, m_{\tilde{g}}, m_{\tilde{u_2}}] +
   2 cos^2 \theta C_{24}^{(1)}+\\
&~&2 sin^2 \theta C_{24}^{(2)}+
   2 p_2 \cdot p_4 cos^2 \theta (C_{12}^{(1)}+ C_{23}^{(1)})+
   2 p_2 \cdot p_4 sin^2 \theta (C_{12}^{(2)} + C_{23}^{(2)})\\
f^{\tilde{u}\tilde{u}\gamma}_{4}=&~&
   - p_2 \cdot p_4 cos^2 \theta (C_{12}^{(1)}+ 2 C_{22}^{(1)})-
     p_2 \cdot p_4 sin^2 \theta (C_{12}^{(2)}+ 2 C_{22}^{(2)})\\
f^{\tilde{u}\tilde{u}\gamma}_{5}=&~&
     cos \theta sin \theta (C_0^{(1)}- C_0^{(2)})+
   2 cos \theta sin \theta (C_{12}^{(1)} - C_{12}^{(2)})\\
\end{eqnarray*}
\begin{eqnarray*}
\{  C_{i}^{(1)},C_{ij}^{(1)}  \}&=&\{ C_{i}^{(1)},C_{ij}^{(1)} \}[-p_3, p_2, m_{\tilde{g}}, m_{\tilde{d_{1}}}, m_{\tilde{d_{1}}}]\\
\{  C_{i}^{(2)},C_{ij}^{(2)}  \}&=&\{ C_{i}^{(2)},C_{ij}^{(2)} \}[-p_3, p_2, m_{\tilde{g}}, m_{\tilde{d_{2}}}, m_{\tilde{d_{2}}}]\\
\end{eqnarray*}
\begin{eqnarray*}
f^{\tilde{d}\tilde{d}\gamma}_{1}=&~&
     sin^2 \theta B_1[m_{d}, m_{\tilde{g}}, m_{\tilde{d_1}}] +
     cos^2 \theta B_1[m_{d}, m_{\tilde{g}}, m_{\tilde{d_2}}] +\\
&~& 2 cos^2 \theta C_{24}^{(1)}+ 2 sin^2 \theta C_{24}^{(2)}\\
f^{\tilde{d}\tilde{d}\gamma}_{2}=&~&
   - sin^2 \theta B_1[m_{d}, m_{\tilde{g}}, m_{\tilde{d_1}}]-
     cos^2 \theta B_1[m_{d}, m_{\tilde{g}}, m_{\tilde{d_2}}]-
   2 cos^2 \theta C_{24}^{(1)}-\\
&~&2 sin^2 \theta C_{24}^{(2)}-
   2 p_2 \cdot p_3  cos^2 \theta (C_{12}^{(1)} + C_{23}^{(1)}) -
   2 p_2 \cdot p_3  sin^2 \theta (C_{12}^{(2)} + C_{23}^{(2)})\\
\end{eqnarray*}
\begin{eqnarray*}
\{ C_{i}^{(1)},C_{ij}^{(1)}  \}&=&\{ C_{i}^{(1)},C_{ij}^{(1)}  \}[-p_4, p_1, m_{\tilde{g}}, m_{\tilde{u_{1}}}, m_{\tilde{d_{1}}}]\\
\{ C_{i}^{(2)},C_{ij}^{(2)}  \}&=&\{ C_{i}^{(2)},C_{ij}^{(2)}  \}[-p_4, p_1, m_{\tilde{g}}, m_{\tilde{u_{1}}}, m_{\tilde{d_{2}}}]\\
\{ C_{i}^{(3)},C_{ij}^{(3)}  \}&=&\{ C_{i}^{(3)},C_{ij}^{(3)}  \}[-p_4, p_1, m_{\tilde{g}}, m_{\tilde{u_{2}}}, m_{\tilde{d_{1}}}]\\
\{ C_{i}^{(4)},C_{ij}^{(4)}  \}&=&\{ C_{i}^{(4)},C_{ij}^{(4)}  \}[-p_4, p_1, m_{\tilde{g}}, m_{\tilde{u_{2}}}, m_{\tilde{d_{2}}}]\\
\end{eqnarray*}
\begin{eqnarray*}
f^{\tilde{u}\tilde{d} w,1}_{1}=&~&
    cos^2 \theta ( B_1[m_{u}, m_{\tilde{g}}, m_{\tilde{u_1}}] +
    B_1[m_{u}, m_{\tilde{g}}, m_{\tilde{d_1}}] )+
    sin^2 \theta (B_1[m_{u}, m_{\tilde{g}}, m_{\tilde{u_2}}] +\\
&~& B_1[m_{u}, m_{\tilde{g}}, m_{\tilde{d_2}}] )+
    4 cos^4 \theta C_{24}^{(1)}+
    4 cos^2 \theta sin^2 \theta (C_{24}^{(2)}+ C_{24}^{(3)})+
    4 sin^4 \theta C_{24}^{(4)}\\
f^{\tilde{u}\tilde{d}w,1}_{2}=&~&
    cos^3 \theta sin \theta  (C_0^{(1)}+ C_0^{(2)})+
    sin^3 \theta cos \theta  (C_0^{(3)}+ C_0^{(4)})+
    \left[ C_0 \rightarrow C_{12} \right]\\
f^{\tilde{u}\tilde{d}w,1}_{3}=&~&
   cos^3 \theta sin \theta (C_{12}^{(1)}+ C_{12}^{(3)})+
   sin^3 \theta cos \theta (C_{12}^{(2)}+ C_{12}^{(4)})+
   \left[ C_{12} \rightarrow C_{23} \right]\\
f^{\tilde{u}\tilde{d}w,1}_{5}=&~&
   cos^3 \theta sin \theta  (C_0^{(1)}+ C_0^{(2)})+
   sin^3 \theta cos \theta  (C_0^{(3)}+ C_0^{(4)})+
   \left[ C_0 \rightarrow C_{11} \right]\\
&~&f^{\tilde{u}\tilde{d}w,1}_{4}= f^{\tilde{u}\tilde{d}w,1}_{1},~~~
f^{\tilde{u}\tilde{d}w,1}_{6}= f^{\tilde{u}\tilde{d}w,1}_{3}\\
\end{eqnarray*}
\begin{eqnarray*}
\{ C_{i}^{(1)},C_{ij}^{(1)} \}&=&\{  C_{i}^{(1)},C_{ij}^{(1)} \}[-p_3, p_1, m_{\tilde{g}}, m_{\tilde{u_{1}}}, m_{\tilde{d_{1}}}]\\
\{ C_{i}^{(2)},C_{ij}^{(2)} \}&=&\{ C_{i}^{(2)},C_{ij}^{(2)} \}[-p_3, p_1, m_{\tilde{g}}, m_{\tilde{u_{1}}}, m_{\tilde{d_{2}}}]\\
\{ C_{i}^{(3)},C_{ij}^{(3)}  \}&=&\{ C_{i}^{(3)},C_{ij}^{(3)} \}[-p_3, p_1, m_{\tilde{g}}, m_{\tilde{u_{2}}}, m_{\tilde{d_{1}}}]\\
\{  C_{i}^{(4)},C_{ij}^{(4)}  \}&=&\{ C_{i}^{(4)},C_{ij}^{(4)} \}[-p_3, p_1, m_{\tilde{g}}, m_{\tilde{u_{2}}}, m_{\tilde{d_{2}}}]\\
\end{eqnarray*}
\begin{eqnarray*}
f^{\tilde{u}\tilde{d}w,2}_{1}=&~&
    cos^2 \theta B_1[m_{u}, m_{\tilde{g}}, m_{\tilde{u_{1}}}]+
    cos^2 \theta B_1[m_{d}, m_{\tilde{g}}, m_{\tilde{d_{1}}}]+
    sin^2 \theta B_1[m_{u}, m_{\tilde{g}}, m_{\tilde{u_{2}}}]+\\
&~& sin^2 \theta B_1[m_{d}, m_{\tilde{g}}, m_{\tilde{d_{2}}}]+
    4 cos^4 \theta C_{24}^{(1)}+
    4 cos^2 \theta sin^2 \theta (C_{24}^{(2)}+C_{24}^{(3)})+
    4 sin^4 \theta C_{24}^{(4)}\\
f^{\tilde{u}\tilde{d}w,2}_{2}=&~&
    cos^3 \theta sin \theta ( C_0^{(1)}+ C_0^{(2)} )+
    sin^3 \theta cos \theta ( C_0^{(3)}+ C_0^{(4)} )+
    \left[ C_0 \rightarrow C_{11} \right]\\
f^{\tilde{u}\tilde{d}w,2}_{5}=&~&
  - cos^3 \theta sin \theta  (C_0^{(1)}+ C_0^{(2)})+
    sin^3 \theta cos \theta  (C_0^{(3)}+ C_0^{(4)})+
    \left[ C_0 \rightarrow C_{11} \right]\\
f^{\tilde{u}\tilde{d}w,2}_{6}=&~&
    cos^4 \theta  C_{12}^{(1)}+
    cos^2 \theta sin^2 \theta (C_{12}^{(2)}+ C_{12}^{(3)})+
    sin^4 \theta  C_{12}^{(4)}+
    \left[ C_0 \rightarrow C_{11} \right]\\
&~&f^{\tilde{u}\tilde{d}w,2}_{3}= f^{\tilde{u}\tilde{d}w,2}_{2},~~~
f^{\tilde{u}\tilde{d}w,2}_{4}= f^{\tilde{u}\tilde{d}w,2}_{1}\\
\end{eqnarray*}
\begin{eqnarray*}
\{ C_{i}^{(1)},C_{ij}^{(1)}  \}&=&\{ C_{i}^{(1)},C_{ij}^{(1)}  \}[-p_4, p_3+p_4, m_{\tilde{g}}, m_{\tilde{u_{1}}}, m_{\tilde{d_{1}}}]\\
\{ C_{i}^{(2)},C_{ij}^{(2)}  \}&=&\{ C_{i}^{(2)},C_{ij}^{(2)}  \}[-p_4, p_3+p_4, m_{\tilde{g}}, m_{\tilde{u_{1}}}, m_{\tilde{d_{2}}}]\\
\{ C_{i}^{(3)},C_{ij}^{(3)}  \}&=&\{ C_{i}^{(3)},C_{ij}^{(3)}  \}[-p_4, p_3+p_4, m_{\tilde{g}}, m_{\tilde{u_{2}}}, m_{\tilde{d_{1}}}]\\
\{ C_{i}^{(4)},C_{ij}^{(4)}  \}&=&\{ C_{i}^{(4)},C_{ij}^{(4)}  \}[-p_4, p_3+p_4, m_{\tilde{g}}, m_{\tilde{u_{2}}}, m_{\tilde{d_{2}}}]\\
\end{eqnarray*}
\begin{eqnarray*}
f^{\tilde{u}\tilde{d}w,3}_{1}&=& \frac{1}{8}(1- \frac{m_W^2}{s})
    [ cos ^{2}\theta (B_1[m_{u}, m_{\tilde{g}}, m_{\tilde{u_1}}]+
    B_1[m_{d}, m_{\tilde{g}}, m_{\tilde{d_1}}])+\\
&~& sin ^{2}\theta (B_1[m_{u}, m_{\tilde{g}}, m_{\tilde{u_2}}]+
    B_1[m_{d}, m_{\tilde{g}}, m_{\tilde{d_2}}]) ] +\\
&~& \frac{cos^4 \theta}{4} [( \frac{m_{W}^4}{s} -m_{W}^2 +
    \frac{2 m_{W}^2}{s} p_1 \cdot p_2 + 2 p_1 \cdot p_4 -
    \frac{2 m_{W}^2}{s} p_1 \cdot p_4 -\\
&~& 2 p_2 \cdot p_4 - \frac{2 m_{W}^2}{s} p_2 \cdot p_4 ) C_{11}^{(1)}+
    2 (  \frac{ m_{W}^4}{s} -  m_{W}^2 +
    \frac{2 m_{W}^2}{s} p_1 \cdot p_2) C_{22}^{(1)}+\\
&~& 2 ( p_1 \cdot p_4 - \frac{ m_{W}^2}{s} p_1 \cdot p_4-
     p_2 \cdot p_4 -  \frac{m_{W}^2}{s} p_2 \cdot p_4 ) C_{23}^{(1)}+
    2 (1 - \frac{m_{W}^2}{s}) C_{24}^{(1)}]\\
&~& sin ^{2}\theta cos ^{2}\theta( [C^{(1)} \rightarrow C^{(2)}]+
    [C^{(1)} \rightarrow C^{(3)}] )+
    sin ^{4}\theta [C^{(1)} \rightarrow C^{(4)}]\\
f^{\tilde{u}\tilde{d}w,3}_{2}&=&
    - \frac{1}{8}(1 + \frac{m_W^2}{s})
    [
    cos ^{2}\theta ( B_1[m_{u}, m_{\tilde{g}}, m_{\tilde{u_1}}]+
    B_1[m_{d}, m_{\tilde{g}}, m_{\tilde{d_1}}] )+\\
&~& sin ^{2}\theta ( B_1[m_{u}, m_{\tilde{g}}, m_{\tilde{u_2}}]+
    B_1[m_{d}, m_{\tilde{g}}, m_{\tilde{d_2}}] )
    ] +\\
&~& \frac{cos^4 \theta}{4}[
  ( \frac{m_{W}^4}{s} -m_{W}^2 +
    \frac{2 m_{W}^2}{s} p_1 \cdot p_2 + 2 p_1 \cdot p_4 -
    \frac{2 m_{W}^2}{s} p_1 \cdot p_4 -\\
&~& 2 p_2 \cdot p_4 - \frac{2 m_{W}^2}{s} p_2 \cdot p_4 ) C_{12}^{(1)}+
    2 (\frac{m_{W}^4}{s} - m_{W}^2 +
     \frac{2 m_{W}^2}{s} p_1 \cdot p_2 ) C_{22}^{(1)}+\\
&~& 2 ( p_1 \cdot p_4 - \frac{ m_{W}^2}{s}  p_1 \cdot p_4 -
      p_2 \cdot p_4 - \frac{ m_{W}^2}{s} p_2 \cdot p_4 ) C_{23}^{(1)} -
   2 (1 + \frac{ m_{W}^2}{s} ) C_{24}^{(1)} ]\\
&~& sin ^{2}\theta cos ^{2}\theta ( [C^{(1)} \rightarrow C^{(2)}]+
    [C^{(1)} \rightarrow C^{(3)}] )+
    sin ^{4}\theta [C^{(1)} \rightarrow C^{(4)}]\\
f^{\tilde{u}\tilde{d}w,3}_{3}=&~&
  - \frac{1}{8} [cos ^{2}\theta ( B_1[m_{u}, m_{\tilde{g}}, m_{\tilde{u_1}}]+
    B_1[m_{d}, m_{\tilde{g}}, m_{\tilde{d_1}}] )+\\
&~& sin ^{2}\theta ( B_1[m_{u}, m_{\tilde{g}}, m_{\tilde{u_2}}]+
    B_1[m_{d}, m_{\tilde{g}}, m_{\tilde{d_2}}] ) ] -\\
&~& cos ^{4}\theta C_{24}^{(1)}-
    sin ^{2}\theta cos ^{2}\theta ( C_{24}^{(2)}+ C_{24}^{(3)} )-
    sin ^{4}\theta C_{24}^{(4)}\\
f^{\tilde{u}\tilde{d}w,3}_{5} =&~&  m_{\tilde{g}}[
    cos ^{3}\theta sin \theta (  C_{11}^{(1)} +C_{0}^{(1)}-
    C_{11}^{(2)} -C_{0}^{(2)})+\\
&~& sin ^{3}\theta cos \theta (  C_{11}^{(3)} +C_{0}^{(3)}-
    C_{11}^{(4)} -C_{0}^{(4)} ) ]\\
f^{\tilde{u}\tilde{d}w,3}_{6}&=&
    cos ^{4}\theta (  C_{12}^{(1)} + C_{23}^{(1)} )+
    cos ^{2}\theta sin ^{2}\theta (  C_{12}^{(2)} +  C_{23}^{(2)}+
    C_{12}^{(3)} +  C_{23}^{(3)} )+\\
&~& 2 sin ^{4}\theta (  C_{12}^{(4)} +  C_{23}^{(4)} ) \\
f^{\tilde{u}\tilde{d}w,3}_{4}&=& - f^{\tilde{u}\tilde{d}w,3}_{3},~~~
f^{\tilde{u}\tilde{d}w,3}_{8}= - f^{\tilde{u}\tilde{d}w,3}_{5},~~~
f^{\tilde{u}\tilde{d}w,3}_{6} = f^{\tilde{u}\tilde{d}w,3}_{7},~~~
f^{\tilde{u}\tilde{d}w,3}_{9} = f^{\tilde{u}\tilde{d}w,3}_{10} = - f^{\tilde{u}\tilde{d}w,3}_{6}\\
\end{eqnarray*}
\begin{eqnarray*}
f^{\tilde{u}\tilde{d} w \gamma }&=&
   m_u cos^4 \theta ( C_{11}^{(1)} - C_{12}^{(1)} ) +
   m_{\tilde{g}} cos^3 \theta sin \theta (C_0^{(2)} - C_0^{(1)}) +
   cos^2 \theta  sin^2 \theta (m_u C_{11}^{(2)} +\\
&~&m_u C_{11}^{(3)} + m_d C_{12}^{(1)} - m_d C_{12}^{(2)}-
   m_u C_{12}^{(2)} - m_d C_{12}^{(3)} -
   m_u C_{12}^{(3)} + m_d C_{12}^{(4)}) +\\
&~&m_{\tilde{g}} cos \theta  sin^3 \theta (C_0^{(4)} - C_0^{(3)}) +
   m_u  sin^4 \theta (C_{11}^{(4)} - C_{12}^{(4)})\\
\end{eqnarray*}
\begin{eqnarray*}
\{ D_{i}^{(1)}, D_{ij}^{(1)}, D_{ijk}^{(1)} \}&=&\{ D_{i}^{(1)}, D_{ij}^{(1)}, D_{ijk}^{(1)} \}[-p_4, p_2, p_1, m_{\tilde{g}}, m_{\tilde{u}_1}, m_{\tilde{u}_1}, m_{\tilde{d}_1}]\\
\{ D_{i}^{(2)}, D_{ij}^{(2)}, D_{ijk}^{(2)} \}&=&\{ D_{i}^{(2)}, D_{ij}^{(2)}, D_{ijk}^{(2)} \}[-p_4, p_2, p_1, m_{\tilde{g}}, m_{\tilde{u}_1}, m_{\tilde{u}_1}, m_{\tilde{d}_2}]\\
\{ D_{i}^{(3)}, D_{ij}^{(3)}, D_{ijk}^{(3)} \}&=&\{ D_{i}^{(3)}, D_{ij}^{(3)}, D_{ijk}^{(3)} \}[-p_4, p_2, p_1, m_{\tilde{g}}, m_{\tilde{u}_2}, m_{\tilde{u}_2}, m_{\tilde{d}_1}]\\
\{ D_{i}^{(4)}, D_{ij}^{(4)}, D_{ijk}^{(4)} \}&=&\{ D_{i}^{(4)}, D_{ij}^{(4)}, D_{ijk}^{(4)} \}[-p_4, p_2, p_1, m_{\tilde{g}}, m_{\tilde{u}_2}, m_{\tilde{u}_2}, m_{\tilde{d}_2}]\\
\end{eqnarray*}
\begin{eqnarray*}
f^{box,1}_{1}&=&
    m_{\tilde{g}} cos^3  \theta  sin \theta D_{27}^{(1)} +
    cos^{2}\theta sin^{2}\theta [D^{(1)} \rightarrow D^{(2)}]-\\
&~& cos^{2}\theta sin^2 \theta [D^{(1)} \rightarrow D^{(3)}]-
    sin^{3}\theta cos \theta [D^{(1)} \rightarrow D^{(4)}]\\
f^{box,1}_{2}&=&
     cos^4 \theta (D_{312}^{(1)} - D_{313}^{(1)}) +
   cos^{2}\theta sin^{2}\theta [D^{(1)} \rightarrow D^{(2)}]+\\
&~&sin^{3}\theta  cos \theta [D^{(1)} \rightarrow D^{(3)}]+
   sin^{3}\theta  cos \theta [D^{(1)} \rightarrow D^{(4)}]\\
f^{box,1}_{3}&=&
     cos^4 \theta (D_{312}^{(1)} - D_{311}^{(1)}) +
   cos^{2}\theta sin^{2}\theta [D^{(1)} \rightarrow D^{(2)}]+\\
&~&sin^{3}\theta  cos \theta [D^{(1)} \rightarrow D^{(3)}]+
   sin^{3}\theta  cos \theta [D^{(1)} \rightarrow D^{(4)}]\\
f^{box,1}_{4}&=&
     cos^4 \theta (D_{313}^{(1)} - D_{311}^{(1)} - D_{27}^{(1)} ) +
   sin^{2}\theta  cos^{2} \theta [D^{(1)} \rightarrow D^{(2)}]+\\
&~&sin^{3}\theta  cos \theta [D^{(1)} \rightarrow D^{(3)}]+
   sin^{3}\theta  cos \theta [D^{(1)} \rightarrow D^{(4)}]\\
f^{box,1}_{5}&=&
    m_{\tilde{g}} cos^3 \theta sin \theta (D_{21}^{(1)} + D_{24}^{(1)}+D_{25}^{(1)}+
   D_{26}^{(1)}+\\
&~&D_{12}^{(1)}-D_{11}^{(1)}) +
   sin^{3}\theta  cos \theta [D^{(1)} \rightarrow D^{(2)}]-\\
&~&sin^{2}\theta  cos^{2} \theta [D^{(1)} \rightarrow D^{(3)}]-
   sin^{3}\theta  cos \theta [D^{(1)} \rightarrow D^{(4)}]\\
f^{box,1}_{6}&=&
    m_{\tilde{g}} cos^4 \theta (D_{22}^{(1)} - D_{24}^{(1)}+D_{25}^{(1)}-
   D_{34}^{(1)}+D_{35}^{(1)}+D_{36}^{(1)}-D_{37}^{(1)}-\\
&~&D_{38}^{(1)}+D_{39}^{(1)}) +
   sin^{2}\theta  cos^{2} \theta [D^{(1)} \rightarrow D^{(2)}]+\\
&~&sin^{3}\theta  cos \theta ([D^{(1)} \rightarrow D^{(3)}]+
   [D^{(1)} \rightarrow D^{(4)}])\\
f^{box,1}_{7}&=&
   cos^4 \theta (D_{27}^{(1)}+D_{311}^{(1)}) +
   sin^{2}\theta  cos^{2} \theta [D^{(1)} \rightarrow D^{(2)}]+\\
&~&sin^{3}\theta  cos \theta [D^{(1)} \rightarrow D^{(3)}]+
   sin^{3}\theta  cos \theta [D^{(1)} \rightarrow D^{(4)}]\\
f^{box,1}_{8}&=&
    m_{\tilde{g}} cos^3 \theta sin \theta (D_{0}^{(1)}+D_{11}^{(1)}+
   -D_{13}^{(1)}+\\
&~& D_{21}^{(1)}-D_{25}^{(1)}) +
   sin^{3}\theta  cos \theta [D^{(1)} \rightarrow D^{(2)}]-\\
&~&sin^{2}\theta  cos^{2} \theta [D^{(1)} \rightarrow D^{(3)}]-
   sin^{3}\theta  cos \theta [D^{(1)} \rightarrow D^{(4)}]\\
f^{box,1}_{9}&=&
   cos^4 \theta (D_{12}^{(1)}-D_{13}^{(1)}+
   D_{23}^{(1)}+D_{24}^{(1)}-D_{25}^{(1)}-D_{26}^{(1)}-D_{310}^{(1)}+D_{34}^{(1)}-\\
&~&D_{35}^{(1)}+D_{37}^{(1)}) +
   cos^{2}\theta  sin^{2} \theta [D^{(1)} \rightarrow D^{(2)}]+\\
&~&sin^{3}\theta  cos \theta [D^{(1)} \rightarrow D^{(3)}]+
   sin^{3}\theta  cos \theta [D^{(1)} \rightarrow D^{(4)}]\\
f^{box,1}_{10}&=&
   cos^4 \theta (D_{27}^{(1)}+D_{311}^{(1)}) +
   cos^{2}\theta  sin^{2} \theta [D^{(1)} \rightarrow D^{(2)}]+\\
&~&sin^{3}\theta  cos \theta [D^{(1)} \rightarrow D^{(3)}]+
   sin^{3}\theta  cos \theta [D^{(1)} \rightarrow D^{(4)}]\\
   f^{box,1}_{11}&=&
    m_{\tilde{g}} cos^3 \theta sin \theta (D_{11}^{(1)}-D_{12}^{(1)}+D_{21}^{(1)}-D_{24}^{(1)}) +
   cos^{2}\theta  sin^{2} \theta [D^{(1)} \rightarrow D^{(2)}]-\\
&~&sin^{2}\theta  cos^{2} \theta [D^{(1)} \rightarrow D^{(3)}]-
   sin^{3}\theta  cos \theta [D^{(1)} \rightarrow D^{(4)}]\\
f^{box,1}_{12}&=&
   cos^4 \theta (D_{24}^{(1)}-D_{22}^{(1)}-
   D_{25}^{(1)}+D_{26}^{(1)}+D_{310}^{(1)}+D_{34}^{(1)}-D_{35}^{(1)}-D_{36}^{(1)}) +\\
&~& sin^{2}\theta  cos^{2} \theta [D^{(1)} \rightarrow D^{(2)}]+
   sin^{3}\theta  cos \theta ([D^{(1)} \rightarrow D^{(3)}]+
   [D^{(1)} \rightarrow D^{(4)}])\\
f^{box,1}_{13}&=&
   - m_{\tilde{g}} cos^3 \theta sin \theta (D_{0}^{(1)}+D_{11}^{(1)}+
   D_{21}^{(1)}) +
   sin^{2}\theta  cos^{2} \theta [D^{(1)} \rightarrow D^{(2)}]-\\
&~&sin^{2}\theta  cos^{2} \theta [D^{(1)} \rightarrow D^{(3)}]-
   sin^{3}\theta  cos \theta [D^{(1)} \rightarrow D^{(4)}]\\
f^{box,1}_{14}&=&
   - cos^4 \theta (D_{13}^{(1)}-D_{12}^{(1)}-2 D_{24}^{(1)}+
   2 D_{25}^{(1)}-D_{34}^{(1)}+D_{35}^{(1)}) +\\
   &~&sin^{2}\theta  cos^{2} \theta [D^{(1)} \rightarrow D^{(2)}]+
   sin^{3}\theta  cos \theta ([D^{(1)} \rightarrow D^{(3)}]+
   [D^{(1)} \rightarrow D^{(4)}])\\
f^{box,2}&=&f^{box,1} ( \{ D_{i}, D_{ij}, D_{ijk} \} [-p_4, p_2, p_1, m_{\tilde{g}}, m_{\tilde{u}_m}, m_{\tilde{u}_m}, m_{\tilde{d}_n}]\\
&~&\longrightarrow \{ D_{i}, D_{ij}, D_{ijk} \}[-p_4, p_1, p_2, m_{\tilde{g}}, m_{\tilde{u}_m}, m_{\tilde{d}_n}, m_{\tilde{d}_n}] )\\
\end{eqnarray*}

\vskip 5mm

Figure  caption:
\vskip 2mm
\noindent
{\bf Figure 1}
The Born diagrams for subprocess $u \bar{d} \longrightarrow W \gamma$.
\vskip 2mm
\noindent
{\bf Figure 2}
The one-loop diagrams in the context of the SUSY QCD for subprocess $u \bar{d}
\longrightarrow W \gamma$.
\vskip 2mm
\noindent
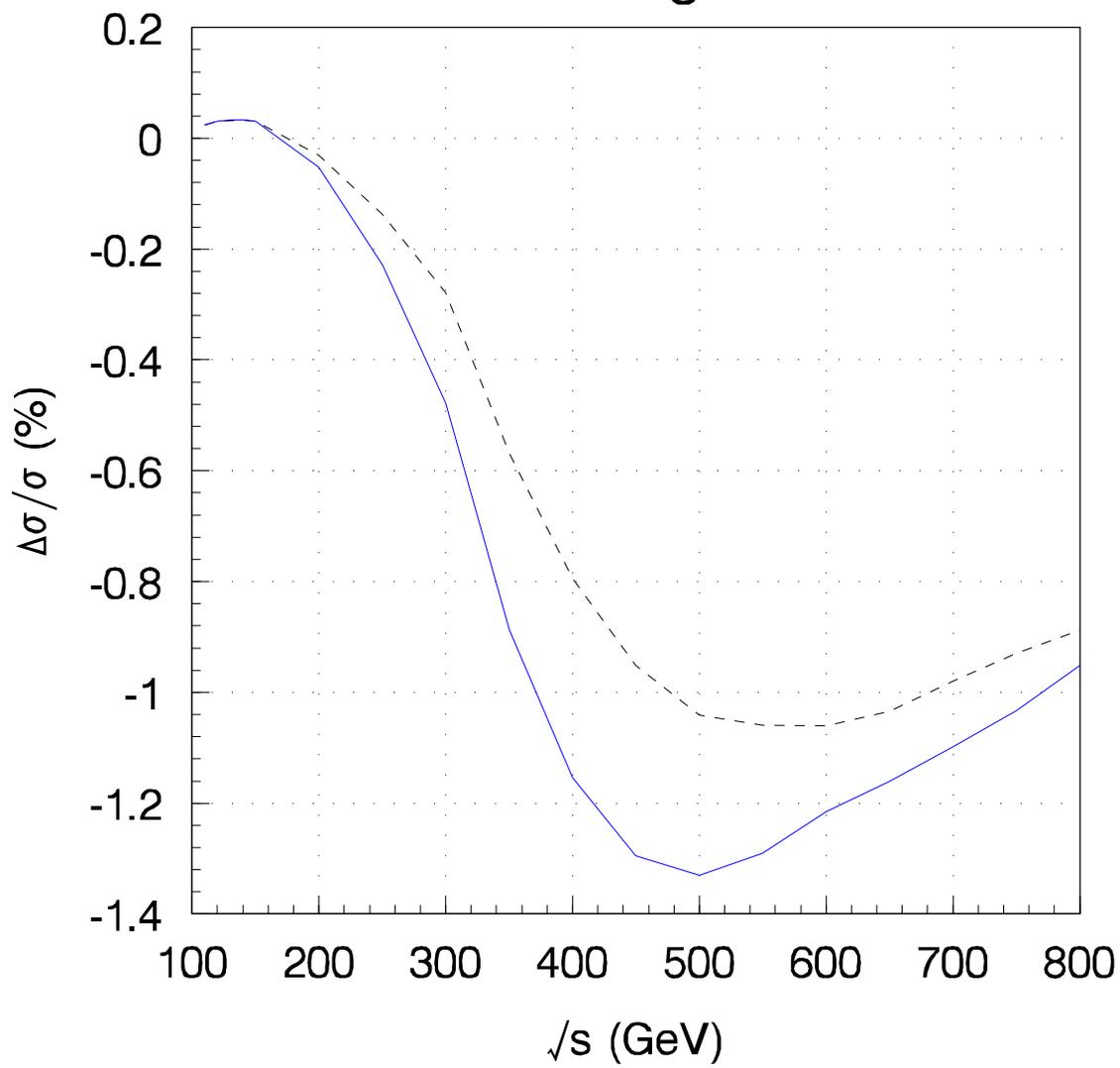
{\bf Figure 3}
The relative corrections of the SUSY QCD for $u \bar{d} \longrightarrow W \gamma$
process as the function of $\sqrt{\hat{s}}$.  The solid line corresponds to
$m_{\tilde{q}}= 175 ~GeV, m_{\tilde{g}}=5 ~GeV $. The dashed line corresponds
to $m_{\tilde{q}}= 175 ~GeV, m_{\tilde{g}}=150 ~GeV $.
\vskip 2mm
\noindent
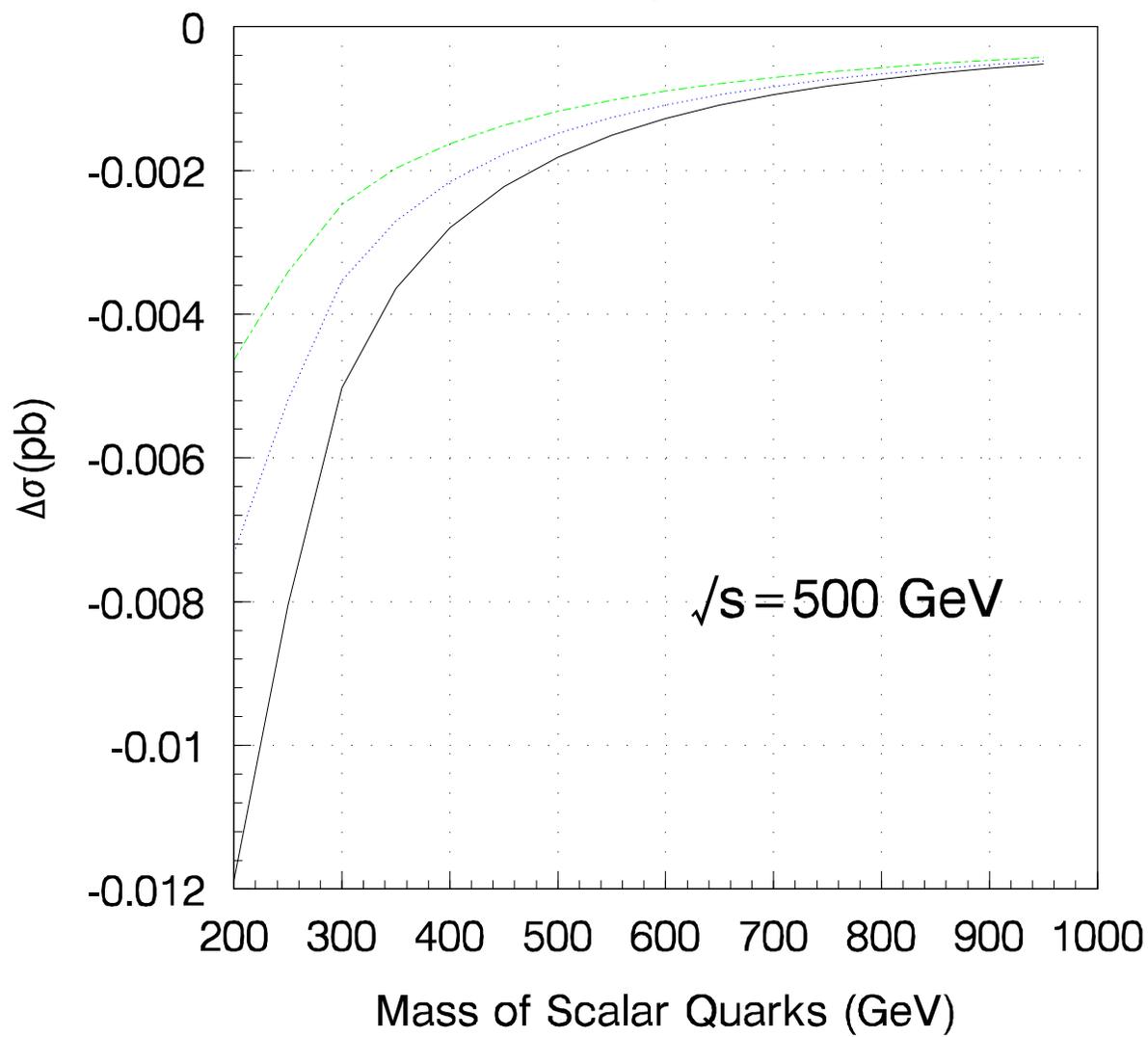
{\bf Figure 4}
The absolute corrections of the SUSY QCD for $u \bar{d} \longrightarrow W \gamma$
process as the function of $m_{\tilde{q}}$ when $\sqrt{s}=500 ~GeV$. The
solid-line, dotted-line and dashed-dotted-line correspond to
$m_{\tilde{g}}= 150 ~GeV, m_{\tilde{g}} = 300 ~GeV$ and
$m_{\tilde{g}}= 500 ~GeV $, respectively.
\vskip 2mm
\noindent
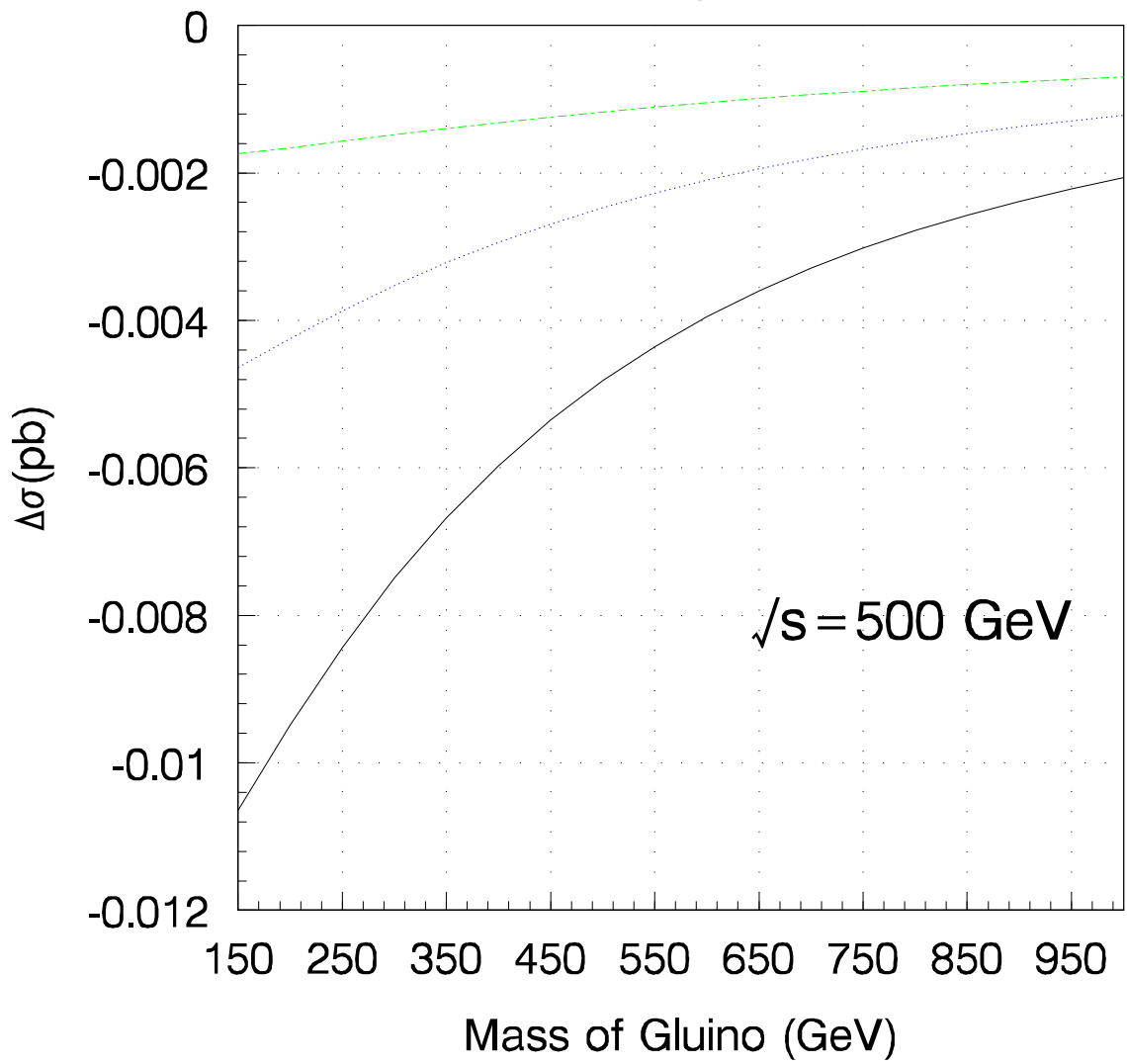
{\bf Figure 5}
The absolute corrections of the SUSY QCD for $u \bar{d} \longrightarrow W \gamma$
process as the function of $m_{\tilde{g}}$ when $\sqrt{s}=500 ~GeV$. The
solid-line, dotted-line and dashed-dotted-line correspond to
$m_{\tilde{q}}= 150 ~GeV$, $m_{\tilde{q}}= 300 ~GeV$ and
$m_{\tilde{q}}= 500 ~GeV$, respectively.
\vskip 2mm
\noindent
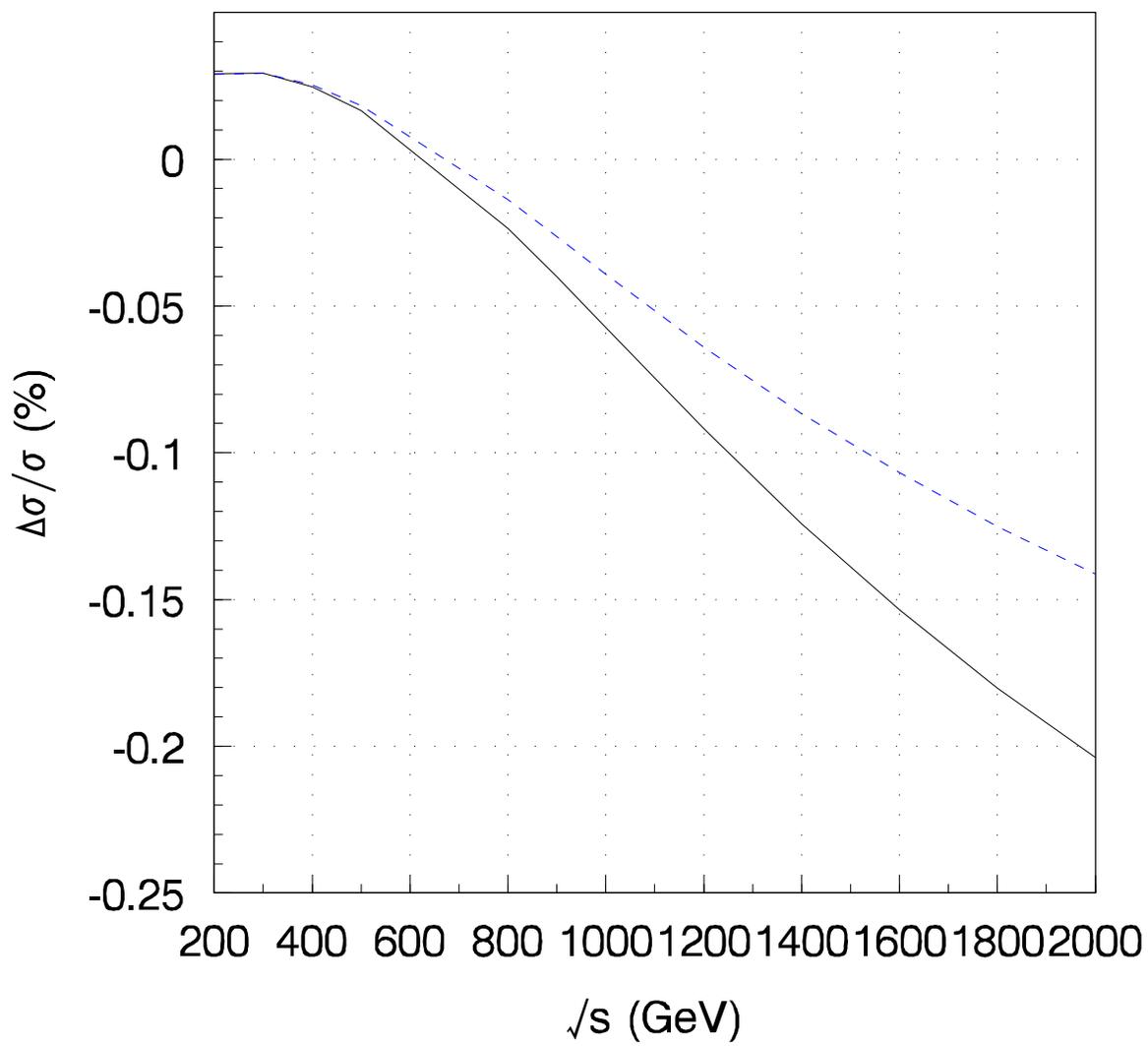
{\bf Figure 6}
The relative corrections of the SUSY QCD for polarized hadronic W $\gamma$
production as the function of $\sqrt{s}$. The polarized parton
distribution functions is adopted from Brodsky's parameterization.
The solid curve corresponds to $p_{L} \bar{p_{L}}$ collision and the dashed
curve corresponds to $p_R \bar{p}_R$ collision with $m_{\tilde{g}}=5 ~GeV$ and
$m_{\tilde{q}}=175 ~GeV$.
\vfil
\eject

\end{large}
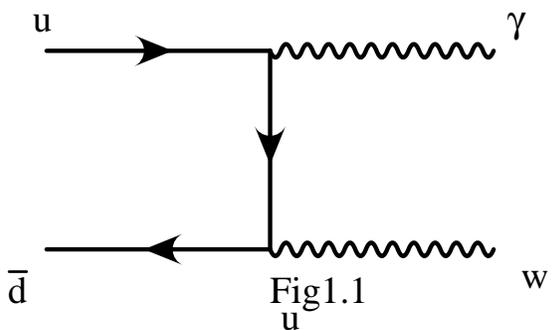
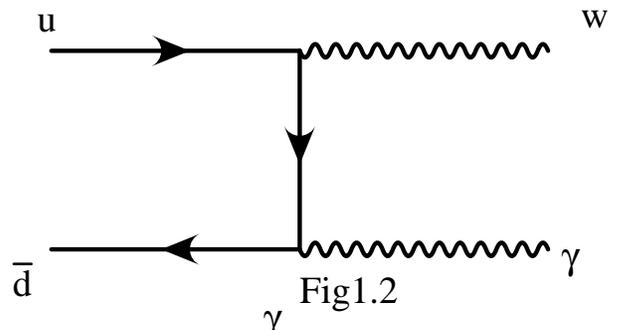
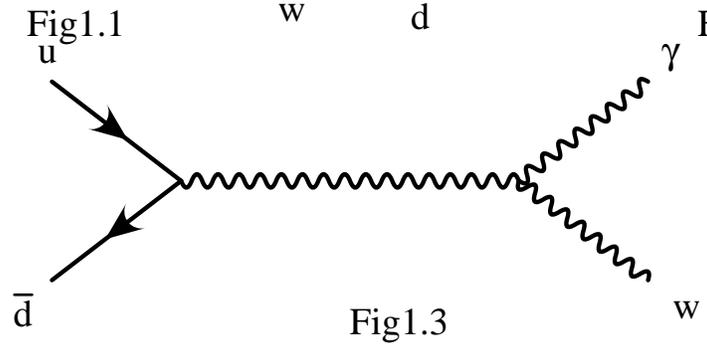
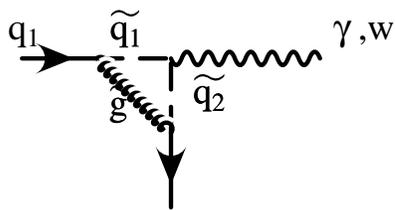
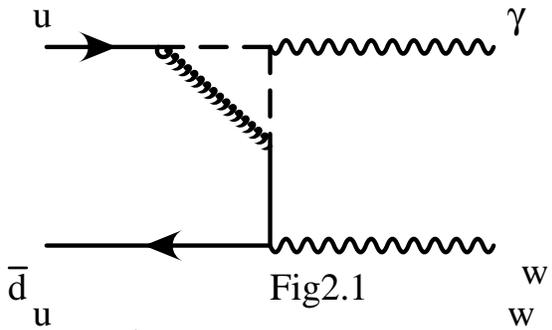
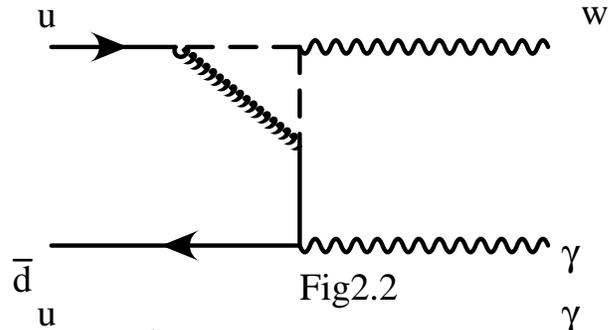
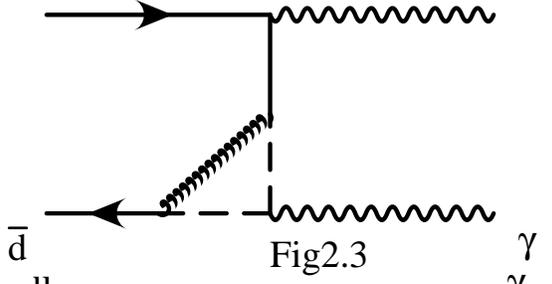
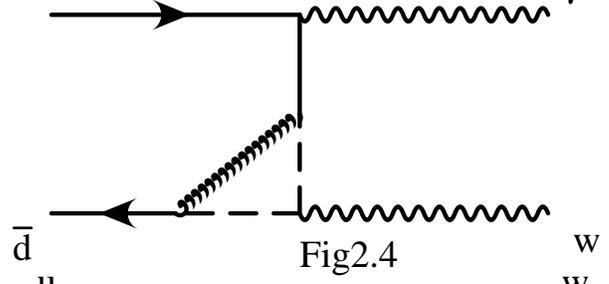
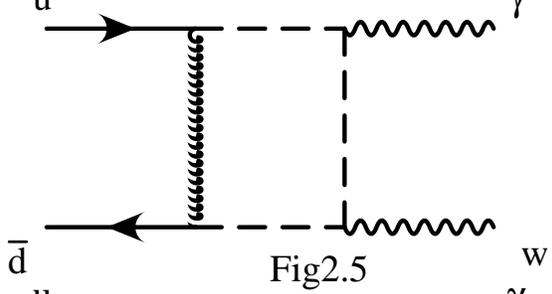
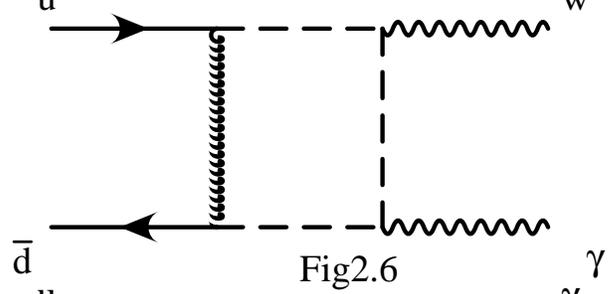
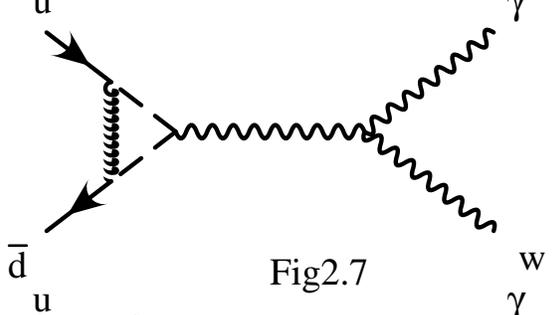
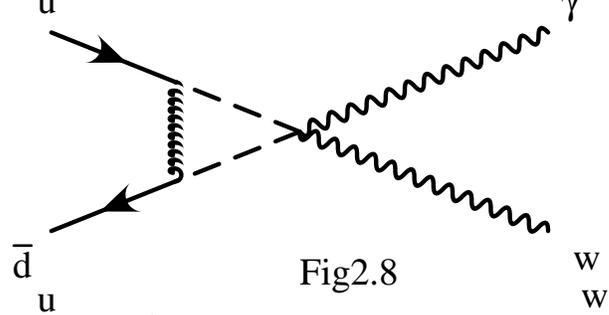
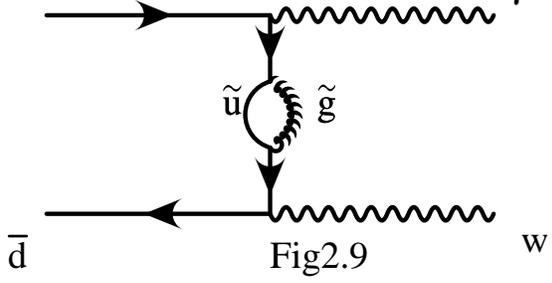
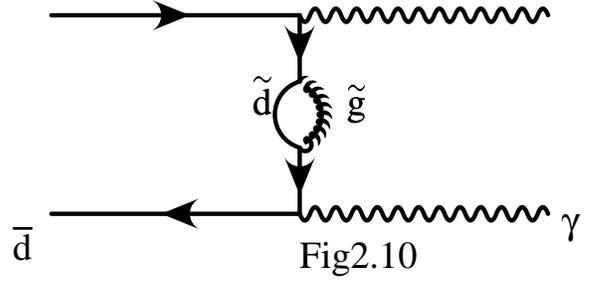
\end{document}